\documentclass{emulateapj}
\usepackage{natbib}
\newcommand{\bvec}[1]{\mbox{\boldmath $#1$}}
\begin{document}
	\title{Double arc instability in the solar corona}
	\author{N. Ishiguro and K. Kusano}
	\affil{Institute for Space-Earth Environmental Research, Nagoya University, Furo-cho, Chikusa-ku, Nagoya, Aichi 464-8601 Japan}
	\email{n-ishiguro@isee.nagoya-u.ac.jp}
	\begin{abstract}
	The stability of the magnetic field in the solar corona 
is important for understanding the causes of solar eruptions.
Although various scenarios have been suggested to date, the tether-cutting reconnection scenario proposed by Moore {\it et al.} (2001) is one of the widely accepted models to explain the onset process of solar eruptions.
Although the tether-cutting reconnection scenario proposed that sigmoidal field formed by the internal reconnection is the magnetic field in pre-eruptive state, the stability of the sigmoidal field has not yet been investigated quantitatively.
In this paper, in order to elucidate the stability problem of pre-eruptive state, we developed a simple numerical analysis, in which the sigmoidal field is modeled by a double arc electric current loop and its stability is analyzed.
As a result, we found that the double arc loop is more easily destabilized than
the axisymmetric torus, and it becomes unstable even if the external field does not decay with altitude, which is in contrast to the axisymmetric torus instability.
This suggests that the tether-cutting reconnection may well work as the onset mechanism of solar eruptions,
and if so the critical condition for eruption under certain geometry may be determined by a new type of instability rather than the torus instability.
Based on them, we propose a new type of instability called double arc instability (DAI). 
We discuss the critical conditions for DAI and derive a new parameter $\kappa$ defined as the product of the magnetic twist and the normalized flux of tether-cutting reconnection.
	\end{abstract}
	\section{INTRODUCTION}
The stability of the current-carrying magnetic flux rope is important for understanding the mechanism of solar eruptions, which are observed as flares, filament eruptions, and coronal mass ejections (CMEs). 
Various studies have tried to elucidate the stability and equilibrium conditions.
Many theoretical models have been developed using the thin current model, 
in which the force acting on the thin electric current loop was calculated to analyze the stability, 
equilibrium conditions, or dynamics of loop, {\it e.g.}, straight current loop model 
\citep{1974A&A....31..189K,1979SoPh...61...89V,1987SvA....31..564M,1991ApJ...373..294F}, 
the axisymmetric current loop model \citep{1989ApJ...338..453C,1999A&A...351..707T,2006PhRvL..96y5002K}, 
and other shape models \citep{1994PhPl....1.3425G,2007ApJ...670.1453I,2013ApJ...771..125O} 
 as reviewed by \citet{2014IAUS..300..184A}.

The axisymmetric torus instability \citep{1966RvPP....2..103S,1978mit..book.....B} was
 applied by \cite{2006PhRvL..96y5002K} to explain the mechanism of solar eruptions using the thin current loop model.
In that study, the flux rope was modeled by a half-circular torus of the electric current rooted on the solar surface. 
An image current is introduced into the sub-photospheric region to satisfy the conditions on the solar surface; hence, the current channel effectively forms an axisymmetric torus. 
In this torus model, only the self-similar expansion of the torus can be allowed; therefore the state variables are the radius $R$ of the torus and the electric current $I$ flowing on the torus.
When an external magnetic field across the torus is imposed, the equilibrium in which the outward hoop force of the torus balances the inward force due to the external field can be satisfied at the condition $I = I_{eq}(R).$

The stability of the torus is determined by the sign of the force acting on the torus when $R$ and $I$ are displaced from the equilibrium state under a constraint such as the conservation of magnetic flux linking the torus. 
For instance, \cite{2006PhRvL..96y5002K} found that when the external field is proportional to $R^{-n}$, the instability requires 
	\begin{equation}
	n > n _{crit} = \frac{3}{2} - \frac{1}{4c} ,
	\label{eq:decay_index}
	\end{equation}
where $n$ is the decay index of the external magnetic field $\bvec{B_{ex}}$, which is defined by
        \begin{equation}
          n \equiv - \frac{\partial \ln{|B_{ex}|} }{\partial \ln{R}} .
          \label{eq:def_decay_index}
        \end{equation}
The coefficient $c$ is given by $c = L / \mu _0 R$, where $L$ is the inductance of the torus, and $\mu_0$ is the permeability of vacuum, respectively.
This result indicates that when the decay index is larger than the critical index $n_{crit}$, the force of the external field becomes weaker than the hoop force if the torus expands from the equilibrium; hence the torus becomes unstable in axisymmetric expansion. 
This is called the torus instability and corresponds to the axisymmetric torus mode of the ideal magnetohydrodynamic (MHD) instabilities in torus plasmas \citep{1966RvPP....2..103S,1978mit..book.....B}. 
\cite{2010ApJ...718.1388D} demonstrated that the loss-of-equilibrium \citep[{$cf.$}][]{1991ApJ...373..294F} and the criticality of torus instability occur at the same physical state. 
\cite{2014ApJ...789...46K} also suggested that the fold catastrophe by loss-of-equilibrium and the torus instability are equivalent descriptions of the onset of solar eruptions.
Although the models of axisymmetric torus instability allow the movement of footpoints across the solar surface, 
\cite{2007ApJ...670.1453I} developed the expression for the line-tied equilibrium of a partial torus, 
based on the flux rope configuration of \cite{1999A&A...351..707T}. 
Numerical stduies confirmed that the torus instability can also occur in the case when the line-tied effect is included and not just in the case of a thin current channel \citep{2007AN....328..743T,2007ApJ...668.1232F}.

Although torus instability can well explain the growth of solar eruptions,  the mechanisms to cause the onset of solar eruption and to initiate the instabilities are still unclear.
Several possible scenarios have been proposed so far.
The injection of magnetic helicity into the flux rope 
owing to the photospheric twist motion possibly
destabilizes the flux rope \citep{1989ApJ...338..453C}, 
which was recently disproved by \cite{2010ApJ...714...68S}.
In addition, the decaying of the external field owing to 
the change of the photospheric magnetic field is another possible 
scenario to access the unstable state \citep{1999ApJ...510..485A,2008ApJ...672.1209B}.
On the other hand, the tether-cutting reconnection scenario proposed by
\cite{2001ApJ...552..833M} is one of the widely cited models 
to explain the formation of an unstable configuration of solar eruptions.
The tether-cutting reconnection scenario explains that the eruption of the sigmoidal field formed by the internal reconnection between sheared field lines can play an important role for the initiation of eruptions.
The ``tether-cutting'' scenario is consistent with the recent observations of \cite{2014ApJ...797L..15C} and the numerical modeling of \cite{2012ApJ...760...31K}, who 
found that the pre-flare reconnection between the sheared arcade and small-scale magnetic flux of typical orientations agrees with the tether-cutting scenario.
The high-resolving recent observation with New Solar Telescope at Big Bear Observatory \citep{2017NatAs...1E..85W} indicates that the detail evolution of pre-flare brightening is well consistent with the flare trigger model by \cite{2012ApJ...760...31K}.

Although it is likely that the instability of the sigmoidal magnetic field causes the onset of solar eruption, the stability of the sigmoidal field is not yet investigated quantitatively due to the complexity of field structure.
The objective of this paper is to shed a light on the stability of the 
sigmoidal field which is formed by the internal reconnection between two sheared fields.
In order to achieve it, we introduce a simple kinematic model, in which the sigmoid is modeled as a double arc shaped electric current loop which is thought to be formed by tether-cutting reconnection, as illustrated in Figure \ref{fig:model_double_arc}(a).
Through the numerical analysis of stability of the simple circuit model
 we also aim to give an answer to the question what determines the critical state of stability of the sigmoidal magnetic field.
 
The paper is organized in following sections.
In section 2, we explain the model for the double arc electric current loop and our numerical analysis. 
We will show the results of the calculations in section 3 and discuss 
the critical parameter of stability of the sigmoidal field based on the numerical 
analysis in section 4. Finally, we summarize the conclusions in section 5.
	\section{MODEL, EQUATIONS, AND NUMERICAL ANALYSIS}
	\subsection{Model}
\begin{figure*}[tbp]
\begin{center}
\includegraphics[width=230mm,bb=0 50 1800 800]{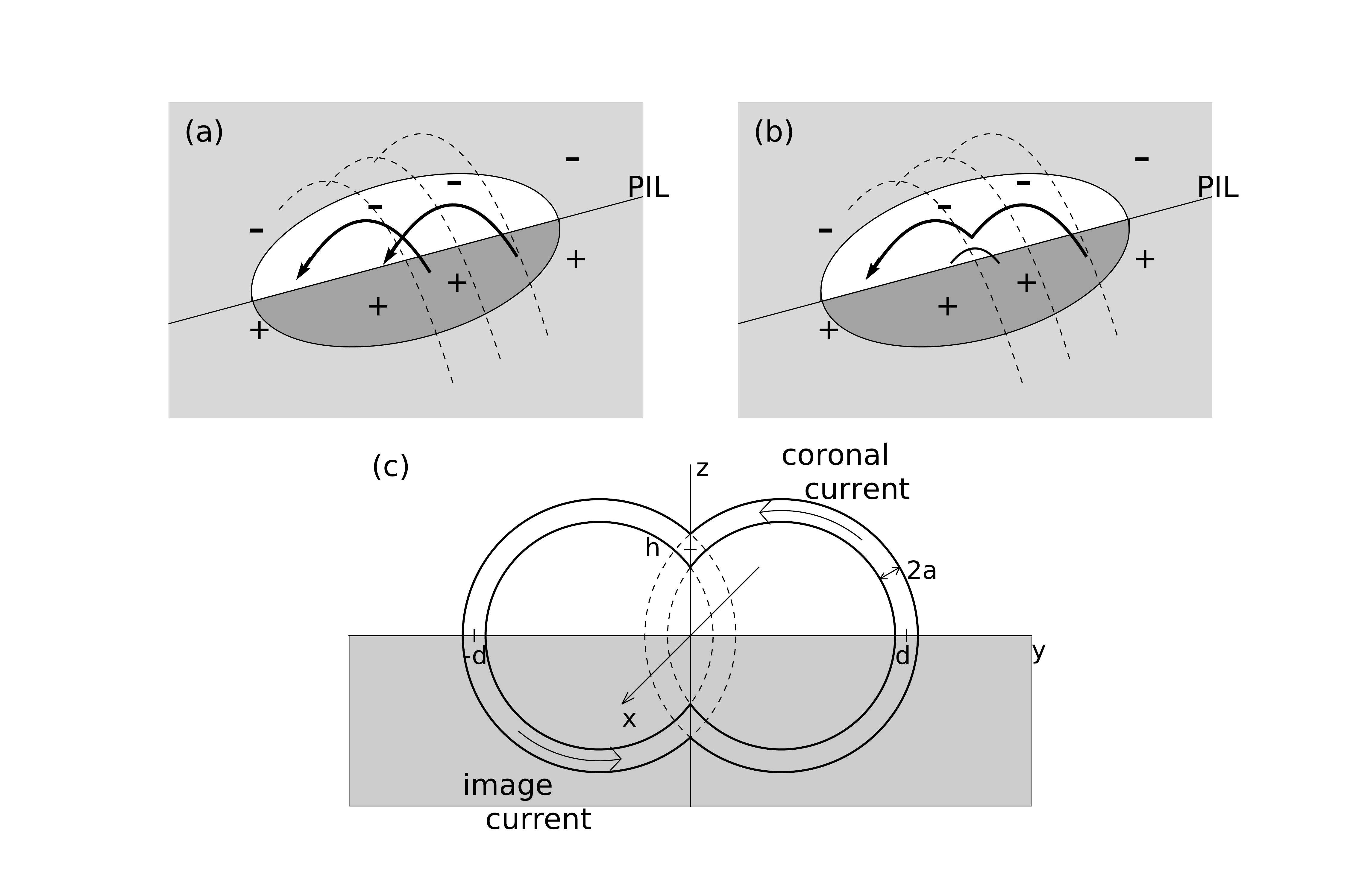}
\end{center}
\caption{(a)(b)The illustration of the tether-cutting reconnection 
in strongly sheared polarity inversion line(PIL).
The dark gray(white) region is positive(negative) polarity region.
The dashed lines are ambient potential magnetic field.
(c)Schematic of the double arc electric current loop.
Solid lines show the shape of the current loop, and the arrows denote the direction of electric current.
The thickness of the loop, $2a$, is uniform. 
The $x$ axis is orthogonal to the plane of loop, that is the $y-z$ plane.
The $x-y$ plane corresponds to the solar surface (photosphere).}
\label{fig:model_double_arc}
\end{figure*}

\begin{figure}[htbp]
\begin{center}
\includegraphics[height=\linewidth*0.5,angle=-90]{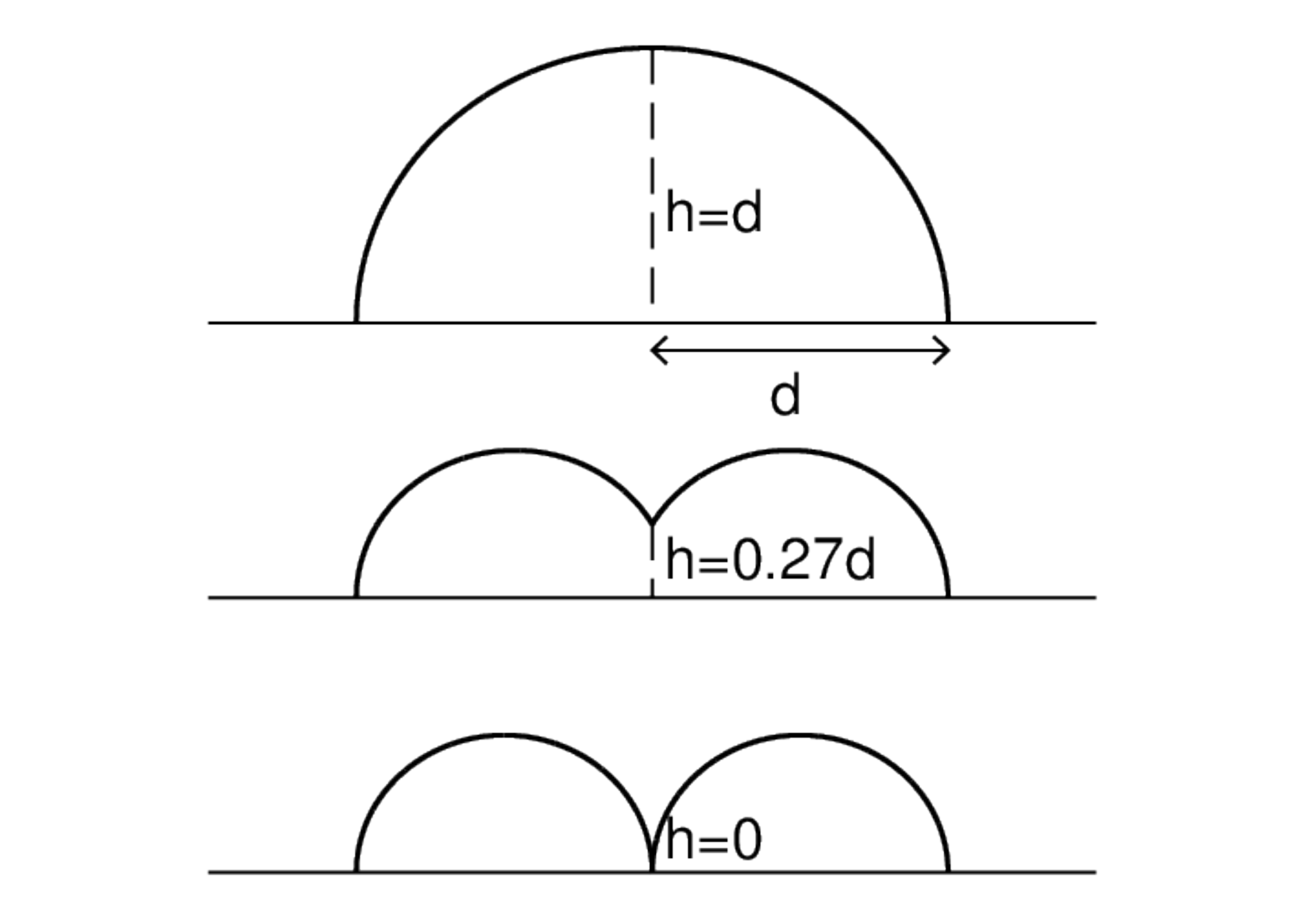}
\end{center}
\caption{Examples of the double arc loop in the coronal region for $h=d, 0.27d$ and $0$. 
The footpoints of the loop are fixed to the solar surface and this constraint corresponds to the line-tied condition. 
The geometry of the loop is parameterized by the joint height $h$.}
\label{fig:ex_loop}
\end{figure}

We model the sigmoidal field as a double arc electric current loop 
(Figure \ref{fig:model_double_arc}b), which can be formed 
by the tether-cutting reconnection as illustrated in Figure \ref{fig:model_double_arc}a.
The double arc electric current loop is assumed to consist of two circles
 which are joined with each other at the center and rooted in the solar surface as shown 
 in Figure \ref{fig:model_double_arc}c.
To simplify the analysis, the double arc loop is assumed to be on the $y-z$ plane,
and the external field $\bvec{B_{ex}}$ on this plane is perpendicular to the plane.
We also assume that the electric current $I$ uniformly flow along the double arc 
and the radius of cross section $a$ is small enough compared with the size of the arc.
The roots of the loop and the joining point of the arcs are at $(y,z)=(\pm d, 0)$ and $(0,h)$, respectively.
Since the roots of the arc are fixed, the state variables are only $h$ and $I$,
and the location of roots $d$ and the external field are fixed. 
The joint height $h$ can vary only in the range from $0$ to $d$, 
in which the double arc becomes a circle of radius $d$ for $h=d$, as shown in Figure \ref{fig:ex_loop}.
It means that this model can be applied only in the early stage of eruption,
and we cannot judge whether the sigmoid can fully erupt to coronal mass ejection with our model.
This model cannot take into consideration the thickness of loop 
and the change in loop shape from circular arc.
These assumptions are introduced because we focus on the onset phase of eruption.
The channel of electric current in the sigmoid may be thin and small 
at the very early stage of tether-cutting reconnection, 
although the sigmoidal field can develop to the thick and big flux rope after the onset of eruption.

In our model, the external magnetic field 
(dotted lines in Figure \ref{fig:model_double_arc}a 
and \ref{fig:model_double_arc}b) is assumed to be a potential field,
and the effects of a magnetic field component pointing along the double arc loop is not taken into account.
Because the tension force of a possible magnetic field component along the double arc loop 
and the Lorentz force due to possible additional electric currents outside the double arc loop,
those are neglected in our study, may work more to destabilize the double arc, 
our study is relevant to the sufficient condition to instability of the double arc loop.

The image current below the solar surface is introduced to satisfy the boundary condition 
that the normal component of the magnetic field on the solar surface is fixed.
Therefore, we analyze ths stability of a figure-8-shaped circuit subject to external magnetic field.
 	
	\subsection{Basic Equations}
Although the current loop under consideration is not axisymmetric, 
we use the zero-dimensional model for the stability analysis, 
which is basically the same as that used for torus instability by \cite{2010ApJ...718.1388D}. 
Here, the term of ``zero-dimensional'' means that the shape of loop is constrained to be circles,
which are characterized only by the parameter $h$, and the stability is analyzed 
by the variation of total energy with respect to the displacement of $h$.
First, we introduce the total magnetic field energy,
	\begin{equation}
		U= \frac{1}{2} L_{tot} I^{2} + I \Phi _{ex},
	\end{equation}
where $L_{tot}$ is the total inductance given by the sum of the external inductance $L_{e}$ and the internal inductance $L_{i}$ of the current loop,
	\begin{equation}
		L_{tot} = L_{e} + L_{i},
		\label{eq:tot_ind}
	\end{equation}
and $\Phi_{ex}$ is the magnetic flux through the area bounded by the double arc current loop and the solar surface. 
The generalized force $F$ conjugate to the coordinate variable $h$ is given by 
the derivative of $U$ with respect to $h$ \citep{1966RvPP....2..103S,1994PhPl....1.3425G}, so that 
    \begin{equation}
		F(h)=\frac{\delta U(h)}{\delta h}=\frac{1}{2} I^{2} \frac{ \partial L_{tot}(h)}{ \partial h} + I \frac{ \partial \Phi _{ex}(h)}{ \partial h}.
		\label{eq:force}
	\end{equation}
From the condition $F=0$, we can obtain the equilibrium current,
	\begin{equation}
		I_{eq}(h) = -2 \frac{\partial \Phi _{ex}(h) / \partial h}{ \partial L_{tot}(h) / \partial h} = -2 \frac{ \partial \Phi _{ex}(h)}{\partial L_{tot}(h)} ,
		\label{eq:eq_cur}
	\end{equation}
as a function of $h$.
Here, note that the generalized force (equation \ref{eq:force}) is not the force acting on each point on the electric current loop, but the net force, because it is derived from the variation of total energy $U(h)$.
So the state corresponding to $F=0$ is the equilibrium under the constraint that the loop is formed by the double arc, but not real equilibrium in which the force on any points should be zero.
We will discuss this limitation also in Section 5.

If the loop evolves under the condition of the ideal magnetohydrodynamics (MHD),
in which any plasma motion is frozen in magnetic field,
the total magnetic flux through the area bounded by the double arc current loop 
and the solar surface must be conserved \citep{2007ApJ...670.1453I,2010ApJ...718.1388D}.
On the other hand, if magnetic reconnection proceeds below the erupting sigmoid, 
the magnetic flux across the loop can increase.
In such a case, because more flux is twisted around the core of sigmoid,
reconnection may work more to destabilize the system.
Therefore, we adopt the constraint of ideal MHD to derive the sufficient condition of instability.

The total magnetic flux $ \Phi _{total}$ is descrived by the following equation.
	\begin{equation}
		\Phi _{total} = L_{e}(h)I + \Phi _{ex}(h) .
	\end{equation}
From this equation, we can derive the evolutional current as a function of $h$,
 	\begin{equation}
		I_{evol}(h) = \frac{1}{L_{e}(h)} \left( \Phi_{total} - \Phi_{ex} (h) \right) ,
		\label{eq:evol_cur}
	\end{equation}
where the conserved flux $\Phi_{total}$ is a parameter to determine the dynamical solution of $h$ and $I$.

In this study, we analyze three different types of external fields, 
$\bvec{B_{ex}^{(i)}}$ for $i=1,2,$ or $3$.
The first type of external field is the potential field
given by the point sources of magnetic flux $\phi$ located at 
$(x,y,z) = (\pm D, 0,0)$, so that they make following magnetic distribution on the $x=0$ plane:
 	\begin{equation}
		\bvec{B_{ex}^{(1)}(x=0,y,z)}= \frac{-4D \phi}{(r^2 + D^2 )^{3/2}} \frac{\bvec{x}}{|\bvec{x}|},
		\label{eq:pole}
	\end{equation}
where $r^2 =y^2 + z^2$.
This field simulates sunspots of bipole type 
as used in the previous study \citep{2010ApJ...718.1388D}.
The second type of external field is the linear force-free field given by the boundary condition of the sinusoidal function of $x$, {\it i.e.} $B_z(x,y,0) = B_0 \sin(x/L)$ where $B_0$ is a constant.
The field on the $x=0$ plane is given by
	\begin{equation}
		\bvec{B_{ex}^{(2)}(x=0,y,z)}=-B _0 e^{-|z|/L} \frac{\bvec{x}}{|\bvec{x}|}.
	\end{equation}
The third type of external field is a uniform field:
	\begin{equation}
		\bvec{B_{ex}^{(3)}(x=0,y,z)} = -B_0 \frac{\bvec{x}}{|\bvec{x}|} .
	\end{equation}
This field corresponds to an extreme case in which the current loop is much smaller 
than the length scale $L$ of the external field.

	\subsection{Numerical Analysis}
Because of the structural complexity of the double arc loop, we use numerical 
analysis to derive the inductance $L_{e}$ and magnetic flux $\Phi_{ex}$,
according to the expression by \cite{1994PhPl....1.3425G}.
The self-flux through the current loop is given by the line integral of the 
vector potential along the inner edge of the loop and the solar surface boundary,
	\begin{equation}
		\Phi = \oint \bvec{A} \cdot d \bvec{r},
		\label{eq:flux_cur}
	\end{equation}
where the vector potential can be obtained by the volume integral of the current loop.
	\begin{equation}
		\bvec{A} = \frac{\mu _0}{4 \pi} \int \frac{I_0}{|\bvec{r}-\bvec{r}'|}d \bvec{r}' .
	\end{equation}
Therefore, the external inductance $L_{e}$ of the loop is calculated as
	\begin{equation}
		L_{e} = {\Phi} / I_{0} = \frac{\mu _0}{4 \pi} \oint \int \frac{1}{|\bvec{r}-\bvec{r}'|}d \bvec{r}' \cdot d \bvec{r} .
		\label{eq:ind_cal}
	\end{equation}
The internal inductance $L_{i}$, which corresponds to the magnetic flux linked to the inner unit current of the loop, is
	\begin{equation}
		L_{i}=\frac{\mu _0 l_l}{8 \pi} ,
		\label{eq:ind_in}
	\end{equation}
where $l_l$ is the length of the loop.
The external magnetic flux $\Phi_{ex}$ is derived by the surface integral over the area bounded by the double arc current loop and the solar surface, 
	\begin{equation}
		\Phi_{ex} = \int \bvec{B_{ex} ^{(i)}} \cdot d \bvec{S} = \oint \bvec{A_{ex} ^{(i)}} \cdot d \bvec{r}.\
		\label{eq:phiex_cal}
	\end{equation}
For this line integration, we use the following vector potentials for the above-mentioned extrnal fields $B_{ex}$:
	\begin{eqnarray}
		\bvec{A_{ex}^{(1)}(x=0,y,z)} &=& \frac{2L \phi}{\sqrt{y^2 + z^2 + D^2}} \left(
		\begin{array}{c}
			0 \\
			\frac{z}{y^2 + D^2} \\
			-\frac{y}{z^2 + D^2}
		\end{array}
		\right),\\
		\bvec{A_{ex}^{(2)}(x=0,y,z)} &=& -B_0 L \frac{z}{|z|} e^{-|z|/L} \frac{\bvec{y}}{|\bvec{y}|},\\
		\bvec{A_{ex}^{(3)}(x=0,y,z)} &=& -B_0 y \frac{\bvec{z}}{|\bvec{z}|} .
		\label{eq:vecpot}
	\end{eqnarray}

We use the second-order central difference to calculate the first-order derivative of Equation $(\ref{eq:eq_cur})$
 and the trapezoidal integration.
The loop is discretized to 16000 grids along its length and 10 $\times$ 30 grids on a cross section (radius $\times$ azimuth) of the loop.

We evaluate the numerical method by comparing the numerical solution of the axisymmetric torus instability with the analytical solution by \cite{2010ApJ...718.1388D} and confirm that the numerical results are consistent with the analytical results for a thin loop, $a \ll d$. Actually, in this study we focus only on the case of a thin loop in which $a/d = 10^{-3}$.

	\section{RESULTS}
	
\begin{figure}[htbp]
\begin{center}
\includegraphics[height=\linewidth*0.5,angle=-90]{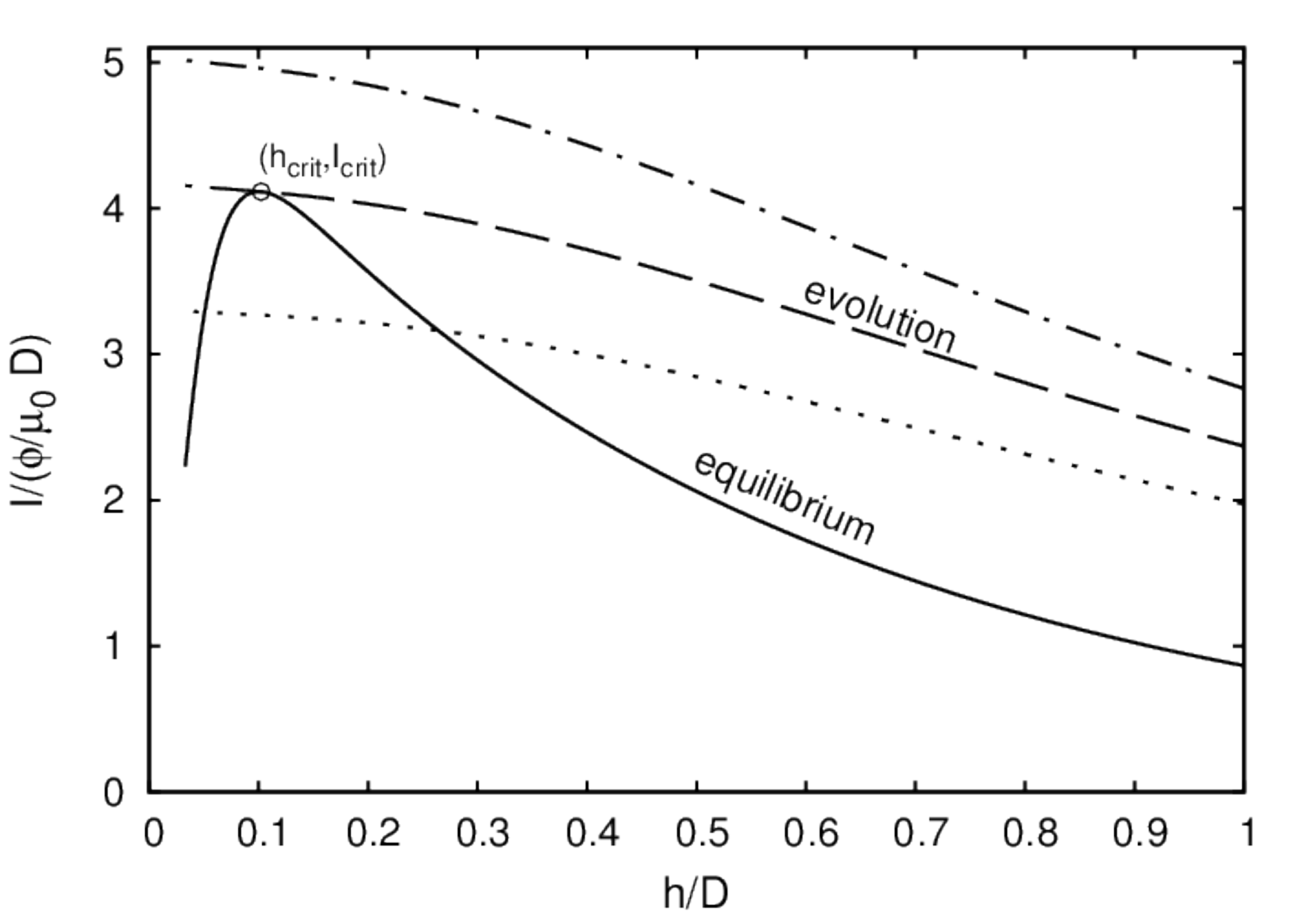}
\end{center}
\caption{The relation between height $h$ and electric current $I$.
The horizontal axis is the normalized joint height h/D, and the vertical axis is the normalized electric current $I/(\phi/\mu_0 D)$. 
The solid line represents the equilibrium state, and the dotted, dashed and dotted-dashed curves correspond to the state for $\Phi_{total}=0.7\Phi_{0}$, $1.0002\Phi_{0}$, and $1.3\Phi_{0}$, respectively.
The circled mark is the critical point, which means that the right branch along the equilibrium curve is unstable and the left is stable.
}
\label{fig:mdi}
\end{figure}
	
In Figure \ref{fig:mdi}, the results for the equilibrium current $I_{eq}$ (solid line) and the evolutional current $I_{evol}$ (dotted, dashed, and dotted-dashed lines) are plotted as a function of $h$ for the first type of external field $\bvec{B_{ex}^{(1)}}$.
The loop below the equilibrium curve moves downward owing to the downward force and is uplifted by the upward force in the region above the equilibrium curve. 
The equilibrium curve has a single peak at $P (h_{p}, I_{p})$ and we define $\Phi_{0} = \Phi_{total} (h_{p}, I_{p})$.
Three evolutional curves are plotted for $\Phi_{total} = 0.7 \Phi_{0}$, $1.0002\Phi_{0}$, and $1.3 \Phi_{0}$, respectively. 
The dashed evolutional curve for $\Phi_{total}=1.0002\Phi_{0}$ is tangent to the equilibrium curve at the point $(h_{crit}, I_{crit})$, where $h_{crit}/D \simeq 0.105$. 
The critical point $(h_{crit}, I_{crit})$ corresponds to the loss-of-equilibrium (LoE) state, above which there is no equilibrium, and the left and right branches of the equilibrium curve from $(h_{crit}, I_{crit})$ are stable and unstable, respectively.
	
\begin{figure}[htbp]
\begin{center}
\includegraphics[height=\linewidth*0.5,angle=-90]{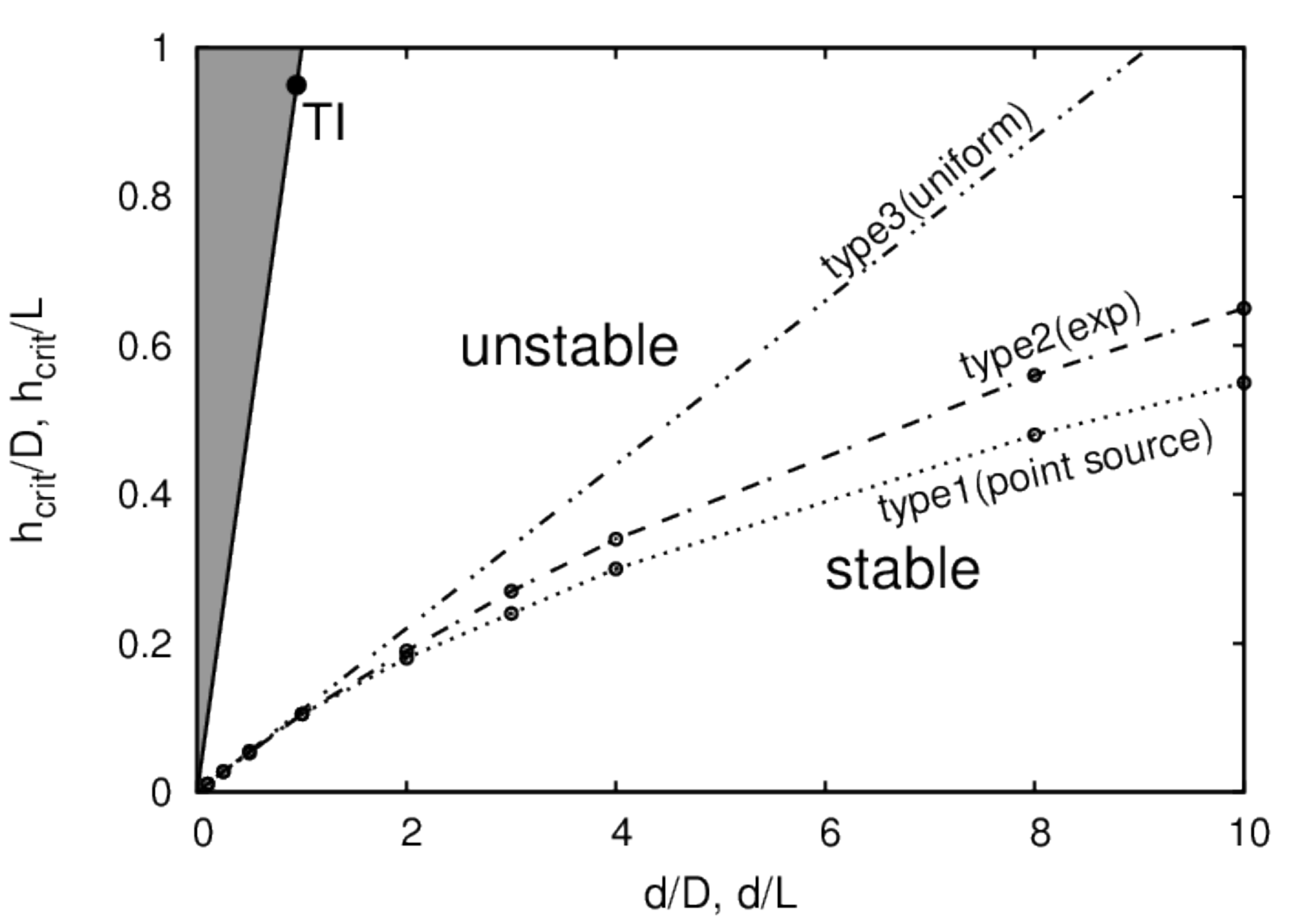}
\end{center}
\caption{The critical height $h_{crit}$ for the type 1 to 3 external fields as a function of the half interval length between footpoints $d$,
which are normalized by $D$ (type 1) or $L$ (type 2). 
The dash-dotted, double-dash-dotted, and dash-double-dotted lines correspond to the types 1, 2, and 3, respectively.
The hatched region is outside the domain defined in the current model $0 \le h \le d$. 
The solid circle marked TI denotes the critical height above which axisymmetric torus instability can grow for the type 1 external field.}
\label{fig:h_vs_R}
\end{figure}

We calculate the critical height $h_{crit}$ of the LoE state for types 2 and 3 of the external fields by the same method and plot the normalized results as a function of $d$ in Figure \ref{fig:h_vs_R}. 

From Figure \ref{fig:h_vs_R}, we find that $h_{crit}$ increases with $d$ in any external field. 
The results for type 1 and 2 are very similar. 
We emphasize that the critical height $h_{crit}$ exists even in the case of uniform external field (type 3). 
The critical height $h_{crit}$ for type 3 is proportional to $d$ because there is no characteristic scale in the external field and the value of it is much smaller than $d$, i.e., $h_{crit} = 0.105d$. 
This suggests that the double arc current loop can be unstable even in the case that the external field does not decay with altitude. 
	
This result is remarkably different from the result of axisymmetric torus instability. 
The axisymmetric loop becomes unstable if and only if the external field decays with the altitude more quickly than a certain rate. 
The condition of torus instability is satisfied when the loop height $h$ and $d$ are larger than $0.95D$ for type 1 external field (corresponding to the solid circle TI in Figure \ref{fig:h_vs_R}). 
The critical height of the double arc loop is much lower than the axisymmetric loop when the external field decays (types 1 and 2). 
This result is consistent with the fact that $h_{crit}$ for types 1 and 2 asymptotically converges to that of type 3 as $d$ tends to zero because the external field in all cases is almost uniform near the bottom surface. 
Therefore, we can conclude that the double arc loop can be more easily destabilized than the axisymmetric loop and that it can become unstable even if the external field does not decay with altitude.
Hereafter, we refer to the instability of the double arc loop as ``double arc instability (DAI)''.
	
The decay index, which is often referred to in the threshold of torus instability, is defined by
	\begin{equation}
		n = -\frac{z}{|B_{ex}|} \frac{\partial |B_{ex}|}{\partial z}.
		\label{eq:def_n}
	\end{equation}
The decay index for type 1 to 3 external fields is given by
	\begin{eqnarray}
		&n^{(1)}& = \frac{3z^2}{y^2 + z^2 + D^2}, \\
		&n^{(2)}& = \frac{|z|}{L}, \\
		&n^{(3)}& = 0.
	\end{eqnarray}

\begin{figure}[htbp]
\begin{center}
\includegraphics[height=\linewidth*0.5,angle=-90]{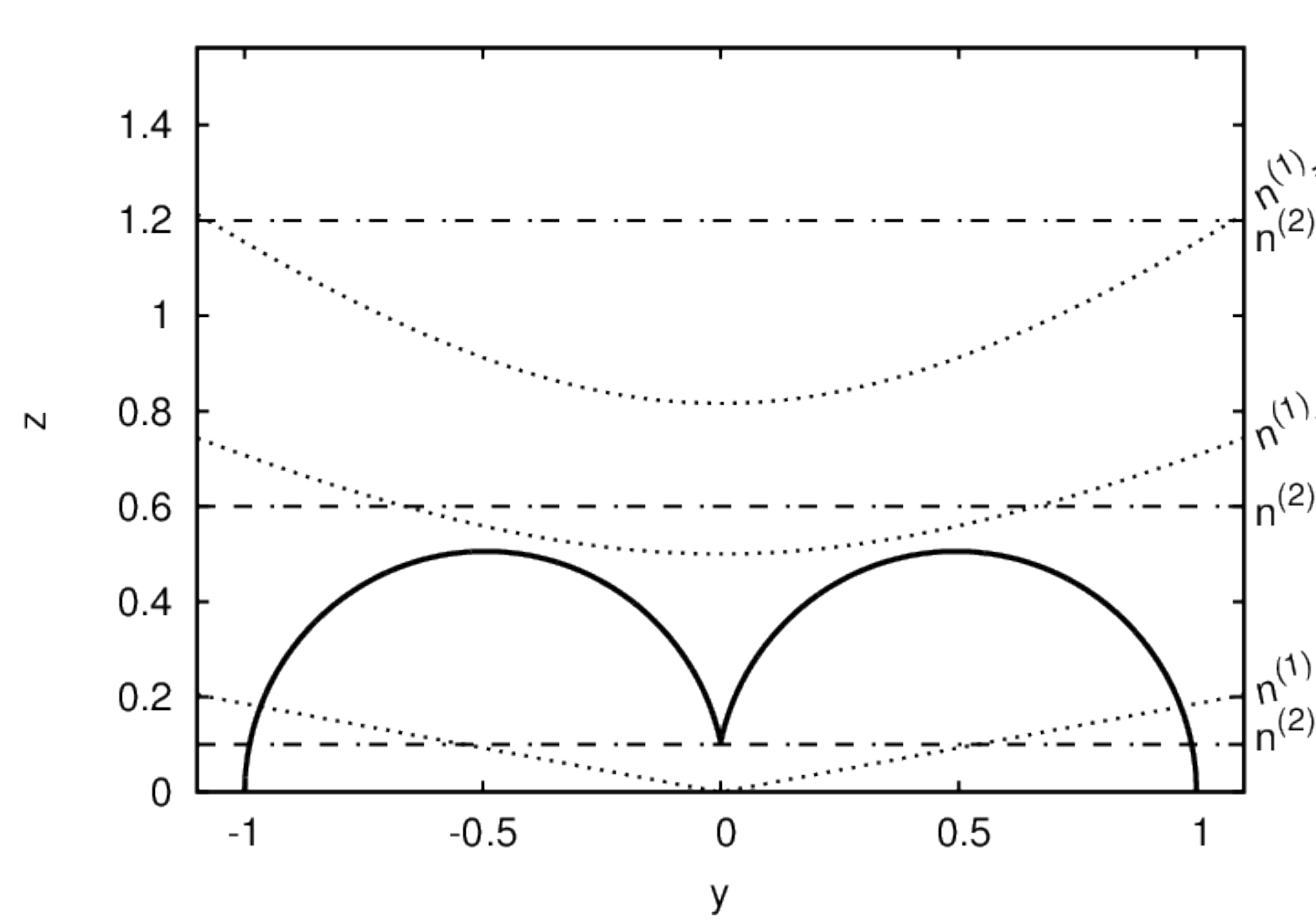}
\end{center}
\caption{The contour map of the decay index for types 1 and 2 on the $y-z$ plane.
Here the $y$ and $z$ axis is normalized by $D$ or $L$, respectively. 
The dashed lines are the contours for $n^{(1)}$ and $n^{(2)} = 0.1, 0.6,$ and $1.2$. 
The solid line shows the double arc loop for $d=1$ and $h=0.105$. }
\label{fig:dec_cont}
\end{figure}
	
Figure \ref{fig:dec_cont} shows the contour map of $n^{(1)}$ and $n^{(2)}$.
The solid curve denotes the double arc loop of the critical state for $d/D=1$ and $h/D=h_{crit}/D$ for the type 1 external field. 
It is clear that the decay index on the loop is below about 0.6, 
which is similar to that in the case of the type 2 external field. 
In contrast, the decay index is zero everywhere in the case of the type 3 external field. 
All the results differ from the axisymmetric torus instability, in which the critical decay index is approximately 1.5 for the type 1 external field \citep{2006PhRvL..96y5002K,2010ApJ...718.1388D}, and it cannot be unstable in the uniform external field (type 3). 
This means that the decay index is not a relevant index for the threshold of the instability for the double arc loop.

\section{DISCUSSION}
In the previous section, we showed that 
the instability of the double arc loop can grow 
even though the torus instability is stable. 
In this section, we discuss the conditions of DAI 
according to the tether-cutting reconnection scenario 
and also the dynamical processes after the onset of instability. 
\subsection{Critical Condition}
\begin{figure}[htbp]
\begin{center}
\includegraphics[height=\linewidth*0.5,angle=-90]{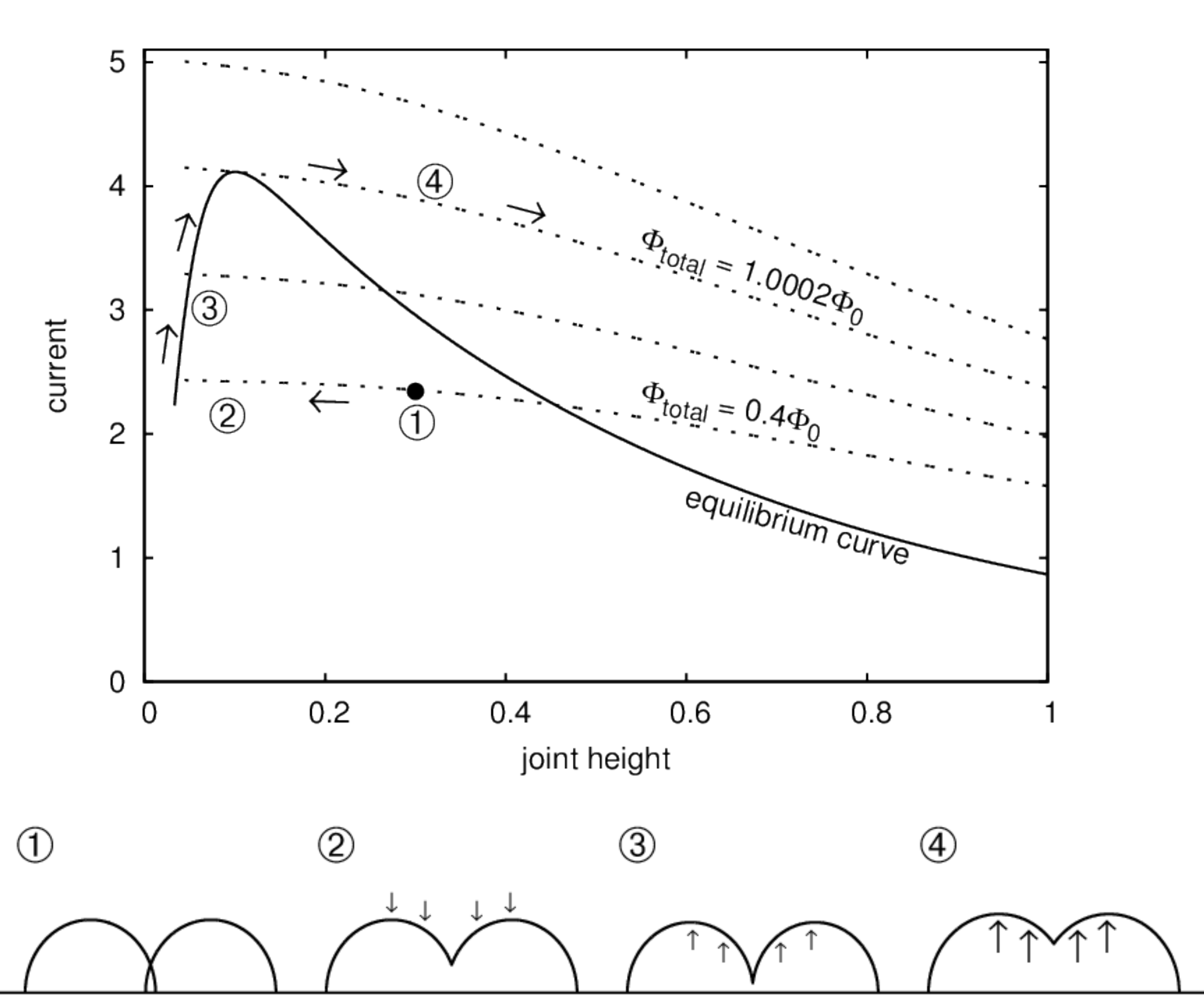}
\end{center}
\caption{The dynamical process of the tether-cutting scenario on the parameter space of the electric current and joint height of the double arc.
The upper diagram is the same as in Figure \ref{fig:mdi}, in which each phase from the tether-cutting reconnection to the onset of eruption, \textcircled{\scriptsize 1} to \textcircled{\scriptsize 4}, is specified.
The lower diagrams show the typical shapes and variations of the double arc in each phase.
}
\label{fig:LoE_ex}
\end{figure}

The tether-cutting scenario proposes that 
the internal reconnection proceeds in the core of 
the sheared magnetic field in the pre-eruptive phase, 
and it may form a double arc flux rope (sigmoidal field) that carries electric current $I$. 
Let us assume that the tether-cutting reconnection proceeds 
at a certain height and forms a double arc loop denoted 
by the solid circle (phase \textcircled{\scriptsize 1}) in Figure \ref{fig:LoE_ex}. 
If the height of tether-cutting reconnection is below the equilibrium curve 
as in Figure \ref{fig:LoE_ex}, the joint height moves down along the evolutional curve 
and reaches the equilibrium state (phase \textcircled{\scriptsize 2}). 
As reconnection further proceeds, current $I$ increases and 
the state variable gradually moves up along the stable 
branch of the equilibrium curve in Figure \ref{fig:LoE_ex} (phase \textcircled{\scriptsize 3}). 
Finally, the loop loses equilibrium and erupts 
when it overcomes the LoE point $(h_{crit}, I_{crit})$ (phase \textcircled{\scriptsize 4}).
Therefore, DAI may control the onset of eruption in the tether-cutting scenario.

Furthermore, we can discuss the critical condition of DAI 
caused by the tether-cutting reconnection based on the result of Figure \ref{fig:LoE_ex}.
Let us assume that the pre-eruptive state is 
approximated by the force-free field, which is represented by 
the force-free equation $\nabla \times \bvec{B} = \alpha \bvec{B}$ 
with the force-free parameter $\alpha$.
Because the electric current density $\bvec{J} = \nabla \times \bvec{B} / \mu_0$ is proportional to $\bvec{B}$ in the 
force-free field, if the magnetic flux $\Phi_{rec}$ is reconnected, 
the current flowing on the double arc loop is proportional to $\Phi_{rec}$, 
{\it i.e.}, 
\begin{equation}
 I = \frac{\alpha}{\mu_0} \Phi_{rec}.
 \label{eq:fff} 
\end{equation}
According to the results of the numerical analysis in Figure \ref{fig:mdi}, 
DAI grows when the electric current $I$ is larger than the critical current
$I_{crit} \simeq 4 \phi / \mu_0 d$ for type 1 external field for $D=d.$ 
This condition $(I>I_{crit})$ corresponds to 
\begin{equation}
 \alpha d > \frac{4\phi}{\Phi_{rec}},
 \label{eq:instability_condition_1}
\end{equation}
owing to the relation of the force-free field $(\ref{eq:fff})$.

On the other hand, the twist of the magnetic field line is defined by
$T=\int \alpha dl / 4 \pi $, where the integration is performed along the
field line from one footpoint to the other.
Because the force free parameter 
$\alpha$ is constant along each field line, the twist of the single arc
of diameter $d$ is given by
\begin{equation}
  T_0 = \frac{1}{4 \pi} \pi \frac{d}{2} \alpha = \frac{\alpha d}{8}.
  \label{eq:twist_arc}
\end{equation}
If we assume the tether-cutting reconnection occurs near footpoint of the field lines of force-free parameter $\alpha$, the twist of double arc is approximated by
\begin{equation}
  T \sim 2 T_0 = \frac{\alpha d} {4}.
\end{equation}
Using this relation and the total magnetic flux of the type 1 external field
$\Phi_{total} = 4 \pi \phi,$
we define a new parameter
\begin{equation}
  \kappa = T \frac{\Phi_{rec}}{\Phi_{total}}.
  \label{eq:def_kappa}
\end{equation}
Then, the condition for instability 
$(\ref{eq:instability_condition_1})$ is rewritten as follows:
\begin{equation}
 \kappa > \frac{1}{4 \pi}
 \label{eq:def_kappa1}
\end{equation}
for the external field of type 1.

In the case of type 2 external field for $L = d$, the critical electric current is given by
\begin{equation}
  I_{crit} \simeq \frac{B_0 d}{\mu_0}.
  \label{eq:I_crit_2}
\end{equation}
If we define the total magnetic flux as
$\Phi_{total} = \int_0^{\infty} dz \int_0^{2d} dy |B_{ex}^{(2)}| = 2d L B_0,$
the critical condition $I>I_{crit}$ is written by
\begin{equation}
  \kappa > \frac{1}{8}
  \label{eq:def_kappa2}
\end{equation}
Also in the case of type 3 external field, the critical electric current is $I_{crit} \simeq 7B_0 d/5 \mu_0$ and, if we define the total flux as
$\Phi_{total} = 2d^2 B_0$, the critical condition is
\begin{equation}
  \kappa > \frac{7}{40}.
  \label{eq:def_kappa3}
\end{equation}

The results above suggest that the critical condition of DAI is in general given by the condition
\begin{equation}
  \kappa > \kappa_0
  \label{eq:kappa_condition}
\end{equation}
in which the threshold $\kappa_0$ depends on the configuration of external magnetic field.
Because $\kappa$ consists of the magnetic twist and the normalized reconnected flux, the critical condition (\ref{eq:kappa_condition}) indicates that the magnetic twist and the tether-cutting reconnection play a complementary role for destabilizing DAI.
If the twist is high enough, even small amount of tether-cutting reconnection may trigger DAI, whereas more reconnection is required in the region of weaker twist.

This result is consistent with the analysis of \cite{2011ApJ...738..161I},
who investigated the twist and connectivity of the magnetic field
of active region NOAA 10930 using vector magnetograms obtained
by the Solar Optical Telescope onboard the Hinode spacecraft and 
the nonlinear force-free field extrapolation.
They showed that the magnetic flux in the flaring region was twisted by more 
than a half turn one day prior to the onset of flares.
If we apply the critical condition (\ref{eq:def_kappa1}) to this active region, since the magnetic twist of field lines prior to tether-cutting reconnection $T_0$ is approximated to be $1/2$, the required flux of tether-cutting reconnection for destabilizing DAI is about $\Phi_{total}/4 \pi$.
\cite{2013ApJ...778...48B} analyzed the magnetic field of this active region using the data measured by Hinode satellite, and found that the magnetic island, which may work for triggering the X3.4 flare at 02:14 UT, 2006 December 13, quickly grew before the onset of the flare.
Though how much fraction of the magnetic island was involved to the tether-cutting reconnection is not clear,
because the area of the magnetic island was as wide as $10\%$ of the major spot prior to the onset of the flare, the result are consistent with the critical condition of DAI.

\cite{2013ApJ...778...48B} also pointed out that another X-class (X1.5) flare occurred at 22:07 UT on 2006 December 14 in the same active region and the size of magnetic island which triggered the event is much smaller than the X-class flare on December 13. The condition of DAI (\ref{eq:kappa_condition}) may provide a possible explanation that the difference of critical size of magnetic island between the two events might be due to the difference of magnetic twist before the two events although we need more investigation to confirm this hypothesis.

The numbers $ \kappa _0 \simeq 0.1 $ to $ 0.2 $ founded in Equation $(\ref{eq:def_kappa1})$ to $(\ref{eq:def_kappa3})$ is also correspond nicely to the analysis with the flux rope insertion method by \cite{2011ApJ...734...53S,2012ApJ...759..105S} and the other studies by their group in 2008 to 2012.
Without the outlier in Figure 8 in \cite{2012ApJ...759..105S}, they calculated the range of flux ratio thresholds $ \Phi _{axi} / \Phi _{tot} \simeq 0.1$ to $ 0.3 $.
Their results are favorable for the thresholds of $\kappa$ if we assume that the magnetic twist is in the range $T \simeq 0.5$ to $1.0$ and the magnitude of $\Phi _{axi}$ corresponds to that of $\Phi _{rec}$.

\subsection{Eruptive Dynamics of DAI}
Finally, we like to discuss the dynamics of instability under the
constraint that the total flux across the double arc loop is
conserved. This corresponds to phase 
\textcircled{\scriptsize 4} in Figure \ref{fig:LoE_ex}. 
Because the generalized force acting on the double arc loop is given by Equation
$(\ref{eq:force})$, the velocity of the joint height $v=dh/dt$ is governed by 
the equation of motion, 
	\begin{equation}
		m \frac{dv}{dt}=\frac{1}{2} I^{2}(h) \frac{ \partial L_{tot}(h)}{ \partial h} + I(h) \frac{ \partial \Phi _{ex}(h)}{ \partial h}, 
		\label{eq:mdi}
	\end{equation}
where, $m$ is the mass of the loop.
	
\begin{figure}[htbp]
\begin{center}
\includegraphics[height=\linewidth*0.5,angle=-90]{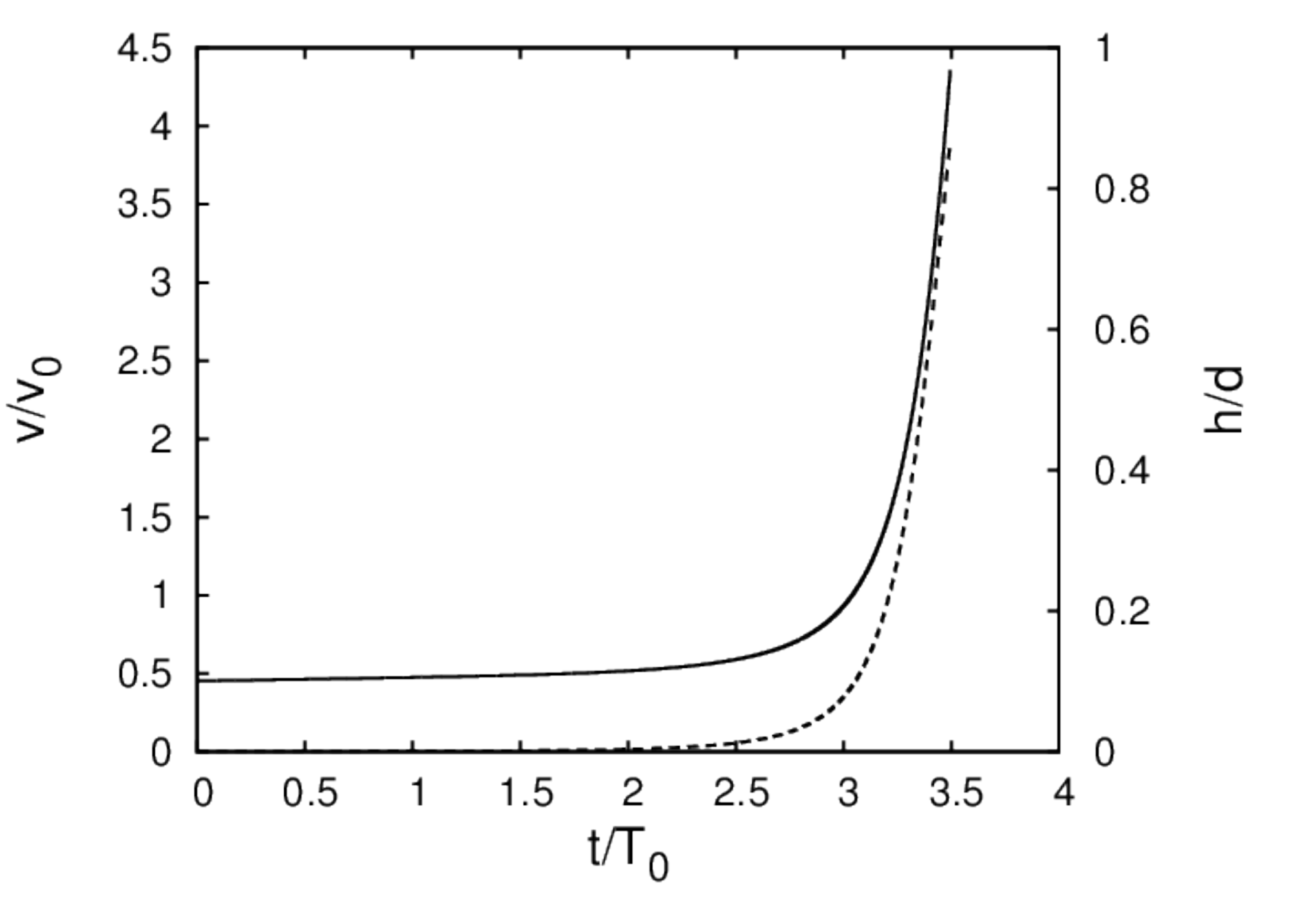}
\end{center}
\caption{Temporal evolution of double arc instability for velocity $v$ (dotted curve) and height of joint point $h$ (solid curve) for the type 1 external field.
  Time $t$ is normalized by the Alfven time $T_0$, and the initial perturbation $ \delta v/v_0 = 0.005$. }
\label{fig:mdihv}
\end{figure}
	
We solve Equation $(\ref{eq:mdi})$ for the critical state by imposing small perturbations. 
Figure \ref{fig:mdihv} shows 
the result of time evolution of the velocity (dotted curve) and
joint height (solid curve) of the loop. 
The joint height $h$, velocity $v$, and time $t$ are normalized 
by $d$, the Alfven speed $v_0 = \phi d^{-2} (\rho \mu _0)^{-1/2} $, and 
the Alfven time $ T_0 = d/v_0 $, respectively, 
where $ \rho $ is the mass density.
The result clearly suggests that while the joint height of the loop slowly increases in the early 
phase before $t/T_0 = 2.5$, 
it quickly erupts afterwards.
The velocity is rapidly 
increases up to $4 v_0$, when the joint height $h$ approaches $d$.

For typical values of the solar corona of 
$d \simeq 40$ Mm, 
$ \rho \simeq 0.5 \times 10^{-9}$ \mbox{kg} \mbox{m$^{-3}$}, and
$ B _0 \simeq 20$ G, 
the factors for normalization are 
$T_{0} \simeq 500$ s and $v_{0} \simeq 80$ \mbox{km} \mbox{s$^{-1}$}. 
Therefore, the results suggest that 
the current loop may accelerate to approximately $320$ \mbox{km} \mbox{s$^{-1}$} when the loop becomes an axisymmetric torus after the slow-rise phase of about 1200 s.
This result is in good agreement with the observations of previous studies \citep[{\it e.g.},][]{2006A&A...458..965C}, which show that the filament accelerated to more than $300$ \mbox{km} \mbox{s$^{-1}$} in about 10 min.
	\section{SUMMARY}
We numerically analyzed the stability of the double arc electric current loop, which may form as a consequence of tether-cutting reconnection and found a new type of instability called DAI. 
Our study suggests that while the critical height of the torus instability depends on the decay index of the external magnetic field \citep{2006PhRvL..96y5002K,2010ApJ...718.1388D}, the critical condition of DAI is insensitive to the decay index.
This is attributed to the fact that DAI is mainly caused by the variation of the inductance (the first term on the right-hand side of Equation (\ref{eq:force})), whereas the torus instability is mainly caused by the variation of the external flux (the second term of Equation (\ref{eq:force})). 
Therefore, the double arc loop can be unstable even in a uniform external field. 
This means that the decay index is not an adequate criterion for the onset of eruption if DAI is responsible to the early phase of solar eruption.
These results reaffirm that the tether-cutting reconnection can efficiently work as the onset mechanism of eruptive events in the solar corona. 

Our study clearly shows that the double arc loop can become unstable irrespective of the decay index, and the unstable double arc loop may obtain substantial kinetic energy when it grows to form an axisymmetric torus. 
Although our model can be applied only to the phase before the double arc loop becomes a torus, it is likely that DAI can play an important role for the acceleration in the early phase of eruption. 
If the double arc loop cannot obtain enough kinetic energy after the growth of DAI and if the loop cannot reach to a region where the decay index does not exceed the critical threshold of the torus instability, the confined eruption will occur.
In other words, however, we cannot judge whether the loop can expand to CME or not only from our model of DAI, because it depends on the interaction of the ejected loop with the magnetic field and plasma in the higher portion. 
Thus, in order to forecast the formation of CME, we have to construct a more generalized model; for instance, by connecting our model to that of the line-tied equilibrium developed by \cite{2007ApJ...670.1453I}. 

Our model is a simple circuit model that cannot describe the rigorous shape of the double arc loop because we do not consider the dynamics of each segment in the loop. 
In particular, it is likely that the cusp shape of the magnetic field line at the tether-cutting reconnection point is quickly relaxed to more smoothed concave shape just after the reconnection, while the shape of each loop is restricted to a circle in our model.
It remians to be solved how the changes in shape of double arc loop affect its stability.
Therefore, we need to develop a more sophisticated numerical simulation based on the MHD equations to verify the dynamics of DAI. 
The development of this type of simulation is currently in progress.
Despite the multiple limitations of our model, the dynamical property of DAI is well consistent with the previous observations and simulations, and we derived the critical condition of DAI, which can be described by the new parameter $\kappa$.
While further study is needed to verify the detailed properties of the DAI, it is likely that the DAI and its critical condition may provide a clue to the understanding of the onset problem of solar eruptions under the specific geometry.
 
\acknowledgments
We wish to thank B. Kliem, S. Imada, S. Inoue, D. Shyukuya and T. Shibayama for their helpful comments and discussions. 
We are grateful to the referee for his/her variable comments to improve the paper.
This work was supported by JSPS/MEXT KAKENHI Grant Nos. JP23340045 and JP15H05814. 
This study was carried by using the computational resource of the Center for Integrated Data Science, Institute for Space-Earth Environmental Research, Nagoya University through the joint research program.
\bibliographystyle{apj}
\bibliography{ms_20170604.bbl}







\end{document}